\documentclass{kapproc} 

\usepackage[dvips]{graphicx}

\newcommand{\fluxunits}{%
   \ensuremath{\mathrm{%
      erg\,s^{-1}}}}

\def\sun{\hbox{$\odot$}}

\def\la{\mathrel{\mathchoice {\vcenter{\offinterlineskip\halign{\hfil
$\displaystyle##$\hfil\cr<\cr\sim\cr}}}
{\vcenter{\offinterlineskip\halign{\hfil$\textstyle##$\hfil\cr
<\cr\sim\cr}}}
{\vcenter{\offinterlineskip\halign{\hfil$\scriptstyle##$\hfil\cr
<\cr\sim\cr}}}
{\vcenter{\offinterlineskip\halign{\hfil$\scriptscriptstyle##$\hfil\cr
<\cr\sim\cr}}}}}

\def\ga{\mathrel{\mathchoice {\vcenter{\offinterlineskip\halign{\hfil
$\displaystyle##$\hfil\cr>\cr\sim\cr}}}
{\vcenter{\offinterlineskip\halign{\hfil$\textstyle##$\hfil\cr
>\cr\sim\cr}}}
{\vcenter{\offinterlineskip\halign{\hfil$\scriptstyle##$\hfil\cr
>\cr\sim\cr}}}
{\vcenter{\offinterlineskip\halign{\hfil$\scriptscriptstyle##$\hfil\cr
>\cr\sim\cr}}}}}

\def\degr{\hbox{$^\circ$}}

\upperandlowercase

\let\footnote\savefootnote

\let\footnotetext\savefootnotetext

\let\footnoterule\savefootnoterule

\kluwerbib 

\begin{document}

\articletitle[]{Massive X-ray binaries}
\articlesubtitle{New developments in the {\it INTEGRAL} era}

\chaptitlerunninghead{Massive X-ray binaries}

\author{Ignacio Negueruela}




\affil{Departamento de F\'{\i}sica, Ingenier\'{\i}a de Sistemas y
  Teor\'{\i}a de la  
Se\~{n}al, Universidad de Alicante, P.O. Box 99, E03080, Alicante,
Spain}
\email{ignacio@dfists.ua.es}

\begin{abstract}
The study of massive X-ray binaries provides important observational
diagnostics for a number of fundamental astrophysical issues, such as the
evolution of massive stars, the stellar winds of massive stars, the formation
of compact objects and accretion processes. More than three
decades of study have led to a coherent picture
of their formation and evolution and some understanding of the
physical mechanisms involved. As more and more systems are discovered,
this picture grows in complexity. Over the last two years, {\it INTEGRAL}
has discovered a new population of massive X-ray binaries,
characterised by absorbed spectra, which challenges some of our
previous assumptions and guarantees that this will be a major subject
of research for the near future.  
\end{abstract}



\section{Introduction}
\label{sec:intro}

Massive X-ray binaries (MXBs) are X-ray sources composed of an
early-type massive star and an accreting compact object (neutron star
or black hole). These are objects of the highest
astrophysical interest, as their study allows us to address a number
of fundamental questions, from the masses of neutron stars to the
  structure of stellar winds (e.g., Kaper 1998). Moreover, because of their
  young age, when considered as a   
  population, they can provide information on properties of galaxies,
  such as their star formation rates (cf.~Grimm et al. 2003). Finally, they
  represent an important phase of massive binary 
  evolution and, again considered as a whole, can provide strong
  constraints on binary evolution and the mechanisms for the formation
  of neutron stars and black holes. A relatively recent list of known
  MXBs is provided by Liu et al. (2000).

The vast majority of MXBs contain X-ray pulsars (magnetised
neutron stars) and can be characterised by the pulse or spin period
$P_{{\rm s}}$. For a few MXBs, X-ray pulsations cannot be detected in
spite of thorough searches. Some of them are considered good black
hole candidates. The 
presence of an X-ray pulsar has allowed the
determination of orbital parameters for a large number of systems via
analysis of Doppler shifts in the pulse arrival times
(e.g., Schreier et al. 1972; Rappaport et al. 1978). When no
pulsations are detected, orbital 
solutions can only be obtained from analysis of radial velocity
changes in the optical components, a method affected by many
uncertainties (cf.~Quaintrell et al. 2003).

For a significant fraction of MXBs, orbital periods ($P_{{\rm orb}}$)
are known through one of the methods described above or because of
a clear periodicity in the X-ray lightcurve. For X-ray pulsars,
Corbet (1986) found that the position of an object in the $P_{{\rm
    orb}}$/$P_{{\rm s}}$ diagram correlates well with other
physical properties. Such a diagram is shown in Fig.~\ref{fig:corbet} for
a number of MXBs. Objects are observed to divide rather cleanly into
three groups.

\begin{itemize}
\item MXBs with short $P_{{\rm s}}$ (a few seconds) and short $P_{{\rm
    orb}}$ (a few days) are observed as very bright
     ($L_{{\rm X}}\sim 10^{38}\fluxunits$) persistent X-ray
    sources. It is believed that matter is being transferred 
    through an accretion disk and that the mass donor is close to
    filling its Roche lobe, though mass is perhaps transferred through
    an enhanced wind. \\
Only a few such systems are known, one in the Milky Way (Cen X-3), one
    in the SMC (SMC X-1) and one in the LMC (LMC X-4). Another system
    in the LMC (LMC X-1) is believed to be the black-hole
    equivalent. These objects are believed to be in a very short-lived
    phase immediately before catastrophic Roche-lobe overflow and so
    to be very rare. However, because of their brightness, these were
    among the first MXBs to be discovered and it is still customary to
    refer to them as ``classical'' MXBs.\\
The mass donors in these systems have luminosity classes around III
    (except SMC X-1, which looks like a low-luminosity
    supergiant). Probably, they have been swollen by the gravitational
    pull of the compact companion.
\item MXBs with long $P_{{\rm s}}$ (hundreds of seconds) and
  relatively short $P_{{\rm orb}}$ are observed as
  moderately bright ($L_{{\rm X}}\sim 10^{36}\fluxunits$)
  persistent X-ray sources with a large degree of variability. The
  optical counterparts are always OB supergiants. The observed
  characteristics are believed to be due to direct accretion from the
  wind of the supergiant on to the neutron star, without a stable
  accretion disk.\\
There are 10 of these systems in the Galaxy, with a couple of
  unconfirmed further candidates. There is one candidate in the LMC
  and none in the SMC. Cyg X-1 would be the black hole
  equivalent. Because of their counterparts, these objects are called
  Supergiant X-ray Binaries (SXBs).
\item MXBs with Be counterparts have relatively long $P_{{\rm
orb}}$. They populate a rather narrow region of the  $P_{{\rm
    orb}}$/$P_{{\rm s}}$ diagram, in spite of the fact that their
X-ray lightcurves show a huge diversity of behaviours. A substantial
number of these objects are X-ray transients, showing very low
X-ray fluxes in quiescence ($<10^{33}\fluxunits$; e.g., Campana et
al. 2002) and luminosities close to the Eddington limit for a
neutron star ($\sim10^{38}\fluxunits$) during bright outbursts.
These systems are known as Be/X-ray binaries (BeXs).

\end{itemize}
 
Only a few among the well studied MXBs fail to fit into these
categories: the  microquasar 2E~0236.6+6101, whose 
counterpart is the B0\,Ve star LS~I~$+61^{\circ}$303
(Massi et al. 2004), the microquasar RX~J1826.2$-$1450, whose
counterpart is the O6.5\,V((f)) star LS~5039 (Rib\'o et al. 2002), and
the peculiar 
wind accretor 4U~2206+54, identified with the peculiar O-type star
BD~$+53^{\circ}$2790 (Negueruela \& Reig 2001). 

\begin{centering}
\begin{figure}[p]
\rotatebox{90}{\vbox to\textwidth{
\hsize=\textheight
\centerline{\includegraphics[bb= 100 150 385 670,angle=-90]{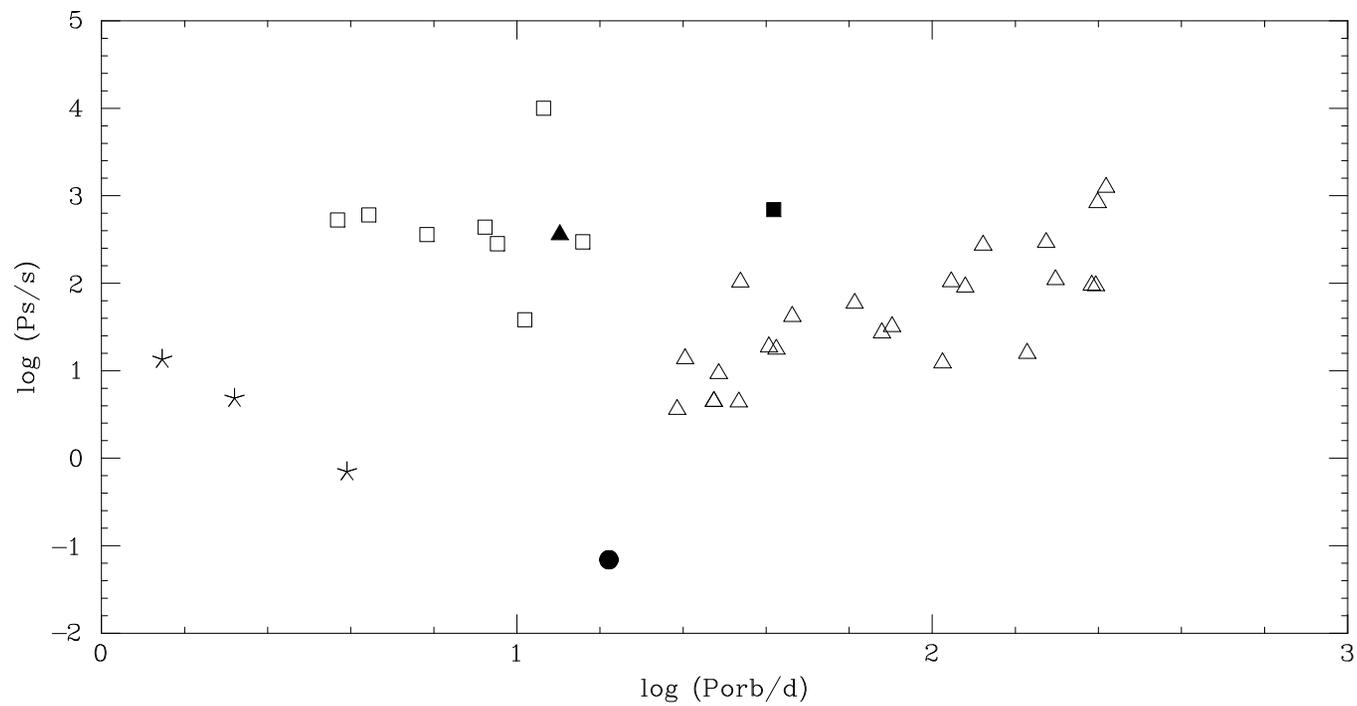}}
\vfill
\caption{The  $P_{{\rm orb}}$/$P_{{\rm s}}$ plot, a modern version of the
  diagram from Corbet (1986). Classical massive X-ray binaries are
  represented as stars, BeX systems are triangles and supergiant
  systems are squares. The filled square is the hypergiant binary
  GX~301$-$02. The filled triangle is SAX~J2103.5+4545. The filled
  circle is the peculiar transient A~0538$-$66. See the text for discussion.}}} 
\end{figure}
\label{fig:corbet}
\end{centering}

Another peculiar system is XTE J0421+560, whose optical counterpart is
the B[e] star CI~Cam. Though B[e] stars are a motley grouping of
objects in very different evolutionary stages, several lines of
argument indicate that CI~Cam is a supergiant (Hynes et al. 2002, and
references therein). XTE~J0421+560 would then be a MXB, even though a
very peculiar one.

\section{The Be/X-ray binaries}
\label{sec:bex}

The class of BeX binaries is generally simply defined by the fact that
the mass donor is a Be star, i.e., a relatively massive star
(sometimes) presenting emission lines because it is surrounded by a
disk of material expelled from the star itself (see Porter \& Rivinius
2003, for a recent review). In this general sense, the microquasar
2E~0236.6+6101 could also be a BeX (see Negueruela 2005, for a general
discussion of the definition). However, BeXs defined in this way
display such a huge variety of properties that one could doubt the
physical reality of this grouping.

Most BeX systems are transients displaying X-ray outbursts and long
periods of quiescence. At present, we believe that we understand the
physical mechanisms leading to this behaviour at the basic level,
though the details are still too complex to allow accurate
predictions. In several systems, we observe relatively quick (a few years)
quasi-periodic cycles, during which the disk around the Be star forms,
grows, gives rise 
to X-ray activity and then disappears (e.g., Negueruela et al. 2001;
Reig et al. 2001). 
Such cycles are governed by a very complex interplay between the
mass loss from the Be star, the dynamics of the Be disk and the
gravitational interaction between the neutron star and the disk
(Okazaki \& Negueruela 2001).

A smaller group of BeX displays much less X-ray variability.
Based on a sample of four pulsars (listed in the upper panel of
Table~\ref{tab:slow}), Reig \& Roche (1999) propose the following
characteristics differentiating ``persistent'' BeX binaries from BeX
transients: 

\begin{enumerate}
\renewcommand{\theenumi}{(\arabic{enumi})}
\item Persistent systems always display relatively long pulse periods,
$P_{{\rm s}}>200\:$s, while most transients have shorter periods.
\item Persistent sources always display low X-ray emission
($\la10^{35}$~erg~s$^{-1}$) with small smooth fluctuations, compared
to the outbursts of transients.
\item The X-ray spectra of persistent sources presents a low cut-off
energy ($\sim4\:{\rm keV}$, compared to $10-20\:{\rm keV}$ in
transients) and no dependence on intensity. 
\item The X-ray spectra of persistent sources does not show the  iron
line at 6.4~keV that is generally seen in transients, or shows it very
weakly.
\end{enumerate}

\begin{table}[!t]\centering
  \setlength{\tabcolsep}{1.2\tabcolsep}
  \caption{
Persistent BeX and candidates. The four objects in the top panel
  define the class (after Reig \& Roche 1999). The first two objects
  in the second panel are SMC BeX binaries with slow pulsations, while
  the other three are Galactic objects with slow pulsations and no
  detected counterpart (and so an identification as cataclysmic
  variables is marginally possible).} 
\label{tab:slow}
\begin{tabular}{lr}
\sphline
Object& $P_{\rm s}$ (s) \\
\sphline
X Per &835\\
RX J0146.9+6121 & 1404\\
RX J0440.9+4431 &202\\
RX J1037.5$-$5647 &862\\ 
\sphline
AX J0049.5$-$7323 & 756\\ 
AX J0103$-$722 & 345\\
1SAX J1452.8$-$5949 &437\\
AX J1749.2$-$2725 & 220\\
AX J1700$-$419 & 715\\
\sphline
  \end{tabular}
\end{table}

Other X-ray pulsars have been proposed to be members of
this class (second panel of
Table~\ref{tab:slow}). Among them, AX~J0049.5$-$7323 and
AX~J0103$-$722 are BeX systems in the SMC sharing many of
the defining characteristics. The three Galactic candidates are
apparently persistent X-ray
sources with long pulsations, but they have no
optical counterpart and there is a small chance that they are really 
cataclysmic binaries.

The division between the two groups is not sharp and indeed there
seem to be exceptions to the apparent rules. For instance, the source
X~0726$-$260 behaves as a persistent source, but has $P_{{\rm
s}}=103\:$s (Corbet \& Peele 1997). A fundamental commonality would be
indicated by the fact that both types of BeX systems occupy the same
region of the $P_{{\rm orb}}$/$P_{{\rm s}}$ diagram. As first noticed by
Corbet (1984), there is a clear correlation between
$P_{{\rm s}}$ and  $P_{{\rm orb}}$ for BeX binaries. The correlation
is not tight, and a few systems do not seem to follow it, for example
X~0726$-$260 and 
XTE~J1946+274 (Wilson et al. 2003). However, Majid et al. (2004) have
demonstrated a 
high degree of correlation between the maximum observed $L_{\rm X}$ and
$P_{\rm s}$ for a large sample of SMC, LMC and Milky Way BeX
systems, something that only makes sense if $P_{\rm s}$ is directly
correlated to the size of the orbit. The only persistent BeX binary
for which the orbital period is known, X~Persei, falls right between
the transient BeX systems in the $P_{{\rm orb}}$/$P_{{\rm s}}$
diagram, and, because of the  $L_{\rm X}$/$P_{\rm s}$ correlation
found by Majid et al. (2004), we have good reason to expect all the other
persistent sources to do the same. 

The reasons for this correlation must lie in the fact that the neutron
star is rotating at some kind of equilibrium spin (Corbet 1986). The
issue was investigated by Waters \& van Kerkwijk (1989), who concluded that the
correlation was due to the neutron stars' rotating at the equilibrium spin
period when the corotation velocity at the magnetospheric radius
equals the Keplerian velocity (see also Stella et al. 1986). They modelled
the Be disk as an 
equatorial slow outflow and concluded that accretion of material from
this wind could spin up the neutron star to this equilibrium period.
Against this conclusion, King (1991) doubted that enough angular
momentum could be supplied through this mechanism. 

On the other hand, our current understanding of the Be
disk is very different from the radial wind model used by Waters \&
van Kerkwijk (1989). We now believe that Be disks are quasi-Keplerian and
held by viscosity (Okazaki \& Negueruela 2001; Okazaki 2001). Such
disks can probably  
transfer large amounts of angular momentum to the neutron star through
an accretion disk (see detailed models in Okazaki et al. 2002;
Hayasaki \& Okazaki 2004), and
indeed $P_{\rm s}$ is observed to decrease strongly during BeX
outbursts (e.g., Finger et al. 1999; Wilson et al. 2003), suggesting
effective transfer 
of angular momentum. Therefore it is likely that the correlation
really stems from an equilibrium between the spin up caused by
accretion of material with angular momentum and spin down during
quiescence, perhaps caused by the propeller effect (Stella et al. 1986;
but see also Rappaport et al. 2004). 

Most BeX binaries have moderate or large orbital eccentricities,
$e\ga0.3$. Evolutionary models have shown that these high
eccentricities can only be explained if supernova explosions are not
symmetric, but rather the neutron star receives a velocity {\it kick}
when it is born (Habets 1987; Portegies Zwart 1995; van den Heuvel \&
van Paradijs 1997). However, the prototype persistent BeX system X~Per was
found to have $e=0.11$ and $P_{{\rm
orb}}=250\:$d (Delgado-Mart\'{\i} et al. 2001). Based on this, Pfahl
et al. (2002) argued
that it was possible to define a new subclass of MXBs 
characterised by long ($P_{{\rm
orb}}>30\:$d) orbital periods and low eccentricities. As the
long orbital periods imply that tidal circularisation cannot have
occurred after the supernova explosion that created the neutron star,
the low eccentricities are primordial: the systems must have formed in
a supernova explosion not accompanied by a {\it kick}.

This class would be
defined by X Per and the four low-eccentricity BeX
transients listed in Table~\ref{tab:circular}.  As can be seen, the
class (from now on, low-$e$ BeX) cuts across the
distinction between persistent and transient BeX, which is based on a
different set of observational parameters. 
There are, however, strong reasons to believe 
that many BeX binaries for which there is not a good orbital
determination may have low $e$, as the model by Okazaki \& Negueruela
(2001) suggests 
that the transient behaviour is intimately related to having a
relatively high $e$. This suggests that all persistent BeX systems,
and also some occasional transients (such as A\,1118$-$615, $P_{{\rm
s}} = 407\:$s), have wide orbits and low eccentricities.

\begin{table}[!t]\centering
  \setlength{\tabcolsep}{1.2\tabcolsep}
  \caption{Be/X-ray transients with low eccentricities. The optical
  counterparts to XTE~J1543$-$568 and 2S~1553$-$542 are actually not
  known, but they are believed to be BeX systems because of their
  transient behaviour. }
  \label{tab:circular}
\begin{tabular}{lcc}
\sphline
    Object& $P_{\rm orb}$ (days) & $e$ \\
\sphline
GS 0834$-$430 & $105.80\pm0.40$ &$<0.17$\\
XTE J1543$-$568 &$75.56\pm0.25$ &$<0.03$\\
KS 1947+300 &$40.42\pm0.02$& $0.03$\\
2S 1553$-$542 &$30.60\pm2.20$ &$<0.09$\\
\sphline
     \end{tabular}
\end{table}

A good counterexample, however, is provided by the BeX transient
KS~1947+300, which has $P_{{\rm orb}}=40.4$~d, $P_{{\rm s}}=18.7$~s
and $e=0.03$ (Galloway et al. 2004). This object has shown a very high
level of X-ray
activity in recent years, in spite of its wide, almost circular orbit,
against the predictions of the model by Okazaki \& Negueruela
(2001). Moreover, its 
low $e$, as well as that of XTE~J1543$-$568, is very difficult to
explain within current models for the formation of BeX binaries.

While the low-$e$ orbit of X Persei may
be explained by a supernova explosion without kick (all the
eccentricity induced is due to mass loss), XTE~J1543$-$568 and KS~1947+300
must have been born with
basically {\bf no} eccentricity. 
Tidal circularisation in such wide orbits can
only occur when the star enters the supergiant phase. The optical
counterpart to KS~1947+300 was studied by Negueruela et al. (2003),
who concluded 
that it was likely of spectral type B0\,Ve. A classification spectrum of
this object was obtained on November 21st, 2002, using the 4.2-m WHT
(La Palma, Spain) and the ISIS spectrograph. The spectrum is displayed
in Fig.~\ref{fig:spectrum} and confirms the spectral classification,
which implies a mass of $\simeq 16M_{\sun}$. In order to explain the
extremely low eccentricity of  KS~1947+300, one would have to assume
that the supernova explosion that
gave birth to the neutron star occurred without a kick and that very
little mass (close to none) was lost during it. The alternative
possibility that the kick velocity exactly balanced the impulse due to
mass loss can be ruled out by the existence of a second system,
XTE~J1543$-$568, with a similarly low $e$. Chance coincidences do not
happen twice.

\begin{centering}
\begin{figure}
\includegraphics[bb=150 175 430 665,clip,angle=-90,width=\columnwidth]{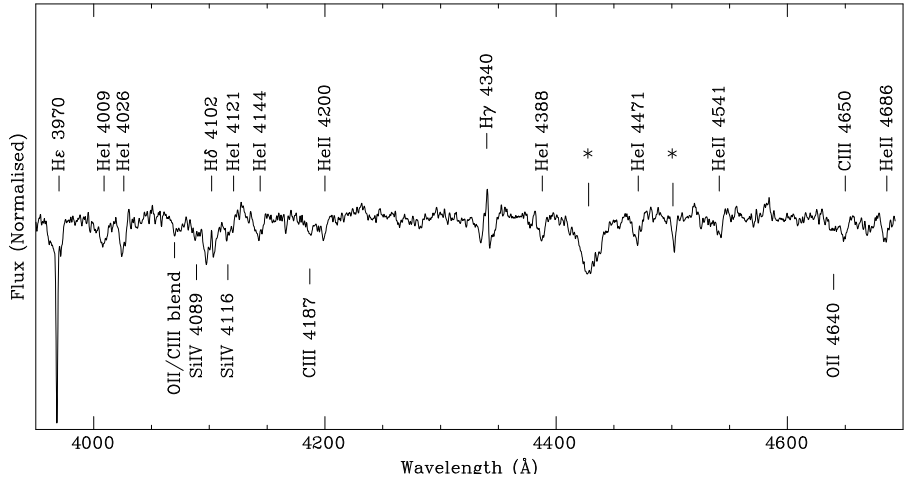}
\caption{Blue spectrum of the optical counterpart to the BeX transient
KS~1947+300, taken on November 21st, 2002, with the 4.2-m WHT + ISIS,
confirming the B0\,Ve spectral type. The parameters of this system are
rather difficult to explain within the current framework. } 
\end{figure}
\label{fig:spectrum}
\end{centering}

Podsiadlowski et al. (2004) have
suggested that the existence of the class of low-$e$ BeX can be
naturally explained by a bimodal distribution in the velocity kicks,
due to different modes of explosion, with neutron stars formed in an
electron-capture supernova receiving a very small kick. Therefore the
class of low-$e$ BeX would be justified by a physical reason and hence
a useful instrument for furthering our
knowledge of MXB formation mechanisms.

\section{The supergiant X-ray binaries}
\label{sec:sxbs}

Because of the evolutionary timescales involved, SXBs are expected to
be much less numerous than BeX binaries. Indeed, the numbers known are
rather small and hence statistical work is unlikely to be very
significant. The mass donors in these systems are OB supergiants, but
they cover a wide range of stellar parameters.

The most massive donor known is probably
HD~153919, the O6.5\,Iaf+ counterpart to 4U~1700$-$37, with an estimated
mass of $58\pm11M_{\sun}$ (Clark et al. 2002). In comparison, the mass
donor in Vela X-1, HD~77581, has spectral type B0.5\,Ib and a mass
in the $23-28M_{\sun}$ range (Quaintrell et al. 2003). This means that
HD~153919 is a much more massive star in a much earlier evolutionary
stage (cf.~Meynet \& Maeder 2003). The X-ray luminosity of a SXB should
depend strongly on the properties of the mass donor, as its
evolutionary status determines the mass loss rate and wind
structure. Because of this, it is a bit surprising that all the
systems known display approximately the same X-ray luminosity,
$L_{{\rm X}}\sim 10^{36}\fluxunits$. 

Many of the SXBs show pulsations with long $P_{{\rm s}}$, generally a
few hundred seconds. In the $P_{{\rm orb}}$/$P_{{\rm s}}$ diagram
(Fig.~\ref{fig:corbet}), they trace 
something close to a horizontal line, as the $P_{{\rm s}}$ does not seem
to be correlated to $P_{{\rm orb}}$. In the standard scenario
(Waters \& van Kerkwijk 1989), it is believed that the neutron stars
were born with 
shorter $P_{{\rm s}}$, but during the main-sequence lifetime of the
present donor (which must have originally been a main-sequence O-type
star), no accretion took place, and the neutron stars were spun down
to these present long $P_{{\rm s}}$ because of the propeller
effect. When the mass donors became supergiants, their mass loss rates
increased sufficiently to permit effective accretion, but this
accretion of matter is not accompanied by transfer of angular
momentum, and so, the pulsars are not spun up back to low values of
$P_{{\rm s}}$. 

This lack of effective angular momentum transfer is probably an
intrinsic quality of the mass transfer mechanism. SXBs display
episodes of both spin up and spin down, that look like random walks
around a particular value (Anzer \& B\"orner 1995). In the classical
Bondi-Hoyle formulation,  accretion
occurs directly from a relatively fast wind and material does not have
enough angular momentum with respect to the accreting body to form a
large and stable accretion disk that can provide the torque to spin
up the neutron star. Early simulations showed the presence of a
``flip-flop'' instability that continuously changed the sign of the
angular momentum accreted (Matsuda et al. 1992).  More realistic simulations
(Ruffert \& Anzer 1995; Ruffert 1999; and 
  references therein) are unable to provide a clear
answer as to how much angular momentum is accreted from a wind.

In Fig.~\ref{fig:corbet}, we can see two systems which deviate
strongly from the straight line traced by all the other SXBs. One is
OAO~1657$-$415, with $P_{{\rm orb}}=10.44\:$d and $P_{{\rm
    s}}=38\:$s (Chakrabarty et al. 1993). This system has displayed long
episodes of spin up, leading to the suggestion that it occupies an
intermediate position between SXBs and classical MXBs. This behaviour
could be perhaps explained if the mass 
donor is a late-B supergiant, as is suggested by the relatively low
mass estimated for it, $M_{*}= 14-18M_{\sun}$ (Chakrabarty et
al. 1993). In this 
case, the supergiant would be on the cool side of the bi-stability
jump (cf.~Vink et al. 2000), and its wind would be rather denser and
slower than those of other 
SXBs, allowing the transfer of angular momentum. Unfortunately, the
counterpart to  OAO~1657$-$415 is very obscured and it has only
recently been detected in the infrared (Chakrabarty et al. 2002).

The second system deviates in the opposite direction. It is 2S~0114+65,
with $P_{{\rm orb}}=11.6\:$d and $P_{{\rm s}}=10000\:$s, the longest
$P_{{\rm s}}$ measured in a neutron star (Hall et al. 2000). The reasons
for this extremely slow pulsation are not clear. It has been suggested
that perhaps the neutron star was born with a very high magnetic field
($B\ga10^{14}\:{\rm G}$, typical of a magnetar) and spun down by the
propeller effect (Li \& van den Heuvel 1999).

There is no obvious correlation between $P_{{\rm orb}}$ and $L_{\rm
  X}$ for SXBs, in contrast to what is observed for BeX systems. As
  noted above, this may simply be due to the number of SXBs being too
  small, as the evolutionary status of the mass donors has to be
  determinant for the $L_{\rm X}$ expected, as well. It is, however,
  curious that no system with $P_{{\rm orb}}>15\:$d is known, as one
  would in principle expect other systems to exist with broader orbits
  and lower X-ray luminosities. The notable exception is GX~301$-$02,
  marked in Fig.~\ref{fig:corbet} as a filled square. This object has
  $P_{{\rm orb}}=41.5\:$d, but it presents two singularities. First,
  its orbit is very eccentric $e=0.46$, meaning that there is a very
  large variation in X-ray flux with the orbital phase (with the maximum
  happening shortly before periastron; cf.~Leahy 2002). Second, the
  optical counterpart is a blue 
  hypergiant, and must thus have a huge radius and a very high mass
  loss rate. As a matter of fact, episodes of rapid spin up have been
  observed in this system (Koh et al. 1997), supporting the idea that
  a transient accretion disk forms. GX~301$-$02 is thus different in
  many senses from other SXBs, reinforcing the idea that general
  properties are difficult to define for this group.

\section{Filling the gaps}
\label{sec:oddballs}

The separation of MXBs into three distinct groups is not only very
clear in the  $P_{{\rm orb}}$/$P_{{\rm s}}$ diagram, but directly
connected to the different physical mechanisms involved in the mass
transfer process and the evolutionary status of the systems.

Among well-studied systems, only the position of the LMC transient
A~0538$-$66 is unclear. This system has sometimes been classed as a BeX
binary, as its optical counterpart is not very luminous (spectral type
around B1\,III; cf.~Negueruela \& Coe 2002) and displays strong H$\alpha$
emission in quiescence. However, A~0538$-$66 is an extremely bright
(super-Eddington) transient X-ray source, displaying optical outbursts
with a $16.6\:$d periodicity (McGowen \& Charles 2003). When the source was
active, X-ray outbursts occurred with the same periodicity,
accompanied by the development of a complex emission spectrum,
characterised by very strong P-Cygni profiles. This lead Charles et
al. (1983)
to propose that the source is likely a very eccentric binary in which
the mass donor occasionally fills its Roche lobe when the compact
object is at periastron. In this sense, as discussed by Corbet (1986),
A~0538$-$66 could be a system evolving toward the classical MXB stage
(via circularisation of its orbit).

The X-ray source SAX J2103.5+4545, discovered by {\it BeppoSAX} in
1997 was found to fall right among the SXBs in the
$P_{{\rm orb}}$/$P_{{\rm s}}$ diagram in spite of being a transient
source (Baykal et al. 2000). Recently, its optical counterpart has
been identified as a 
B0\,Ve mild Be star (Reig et al. 2004a). Its position in the diagram,
marked by a filled triangle in
Fig.~\ref{fig:corbet}, though
certainly surprising, can perhaps be explained within the standard
framework if we assume that the mass donor has only recently become a
Be star or displays very sporadic activity. In this case, the spin
evolution of the neutron star would have carried it to a value typical
for a non-accreting system (as the spin periods of SXBs are believed
to be) and BeX-like activity could not have been able to bring it back
to a shorter spin yet. If this interpretation is correct, one can
suspect that substantial numbers of binaries formed by a normal OB
main-sequence star and a neutron star exist, but they are not
detected, as they do not produce X-ray emission (LS~5039 is, for some
reason, an exception, as the optical star is an O6.5\,V
star; its X-ray luminosity is, however, rather low). If the evolution of
the OB star turns it into a Be star, accretion will start, and it will
move from a $P_{{\rm s}}$ of several hundred seconds to the equilibrium
  position corresponding to its $P_{{\rm orb}}$. If the OB star does
  not display Be characteristics, it will remain hidden until it
  begins to swell and evolve toward the supergiant stage.

Another system recently found that may represent an intermediate stage
in the evolution of MXBs is IGR~J00370+61226. This source, detected
with {\it INTEGRAL} has been identified with 1RXS J003709.6+612131, a
{\it ROSAT} source coincident with the catalogued OB star BD
$+60\degr$73. Observations with {\it RXTE}/ASM reveal a clear
periodicity of $15.665\pm0.006\:$d. The source is hardly detected
during most of the time, and displays weak outbursts with this
periodicity (den Hartog et al. 2004). Optical observations of BD $+60\degr$73
reveal a spectral type BN0.5\,II-III (Negueruela \& Reig 2004; Reig et
al. 2005), and no indications of emission 
lines. Therefore the optical counterpart is not a Be star and has a
luminosity far too low to have a strong stellar wind. The source of
the material accreted is not clear, but a likely hypothesis could be
that a compact object in a very eccentric orbit comes very close to
the surface of the B star. In this case, IGR~J00370+61226 would
represent an evolutionary stage previous to that of
A~0538$-$66. Obviously, the determination of the orbital parameters of
this system would be of high importance.

\section{The new era}

The {\it INTEGRAL} observatory, launched in October 2002,
contains X-ray and $\gamma$-ray telescopes
(see Winkler et al. 2003). For the last two years, it has detected
several new X-ray sources, especially toward the inner regions of the
Galaxy, which are monitored very frequently. Many of these new sources
have rather hard spectra, with little or no emission in the soft
X-rays. This lack of soft X-ray emission is probably the reason why
they have not been detected by previous missions. An up-to-date list
of the new sources is kept by J. Rodriguez at {\tt  
http://isdcul3.unige.ch/~rodrigue/html/igrsources.html}.

The first such system to be detected was IGR J16318$-$4848, which was
observed to be hard because its soft flux was absorbed by intervening
material. Observations with several satellites showed that the
photoelectric absorption was variable and that the amount of absorbing
material was much higher than the interstellar column density in that
direction (Matt \& Guainazzi 2003), leading to the suggestion that it
was an X-ray 
binary immersed in a dense envelope of material associated with the
donor star (Revnivtsev et al. 2003). The search for the counterpart led to the
discovery of a very reddened object, undetectable in optical bands,
but rather strong in near infrared. Infrared spectra of this object
reveal unprecedented characteristics, with a wealth of emission lines
corresponding to both high and low excitation elements
(Filliatre \& Chaty 2004) that has not been observed in any other source
known. There is, however, some similarity to CI~Cam, the optical
counterpart to XTE~J0421+560, leading Filliatre \& Chaty (2004) to the
hypothesis 
that the counterpart may also be a sgB[e] star and that the high
absorption is caused by a thick disk of material surrounding it.

Subsequently, {\it INTEGRAL} has discovered several other sources
characterised by absorbed spectra. Many of them have later been
found in archival observations obtained in the past with other
satellites. Among them, IGR J16320$-$4751 had previously been observed
with {\it ASCA}. Its X-ray spectrum was similar to that of accreting
pulsars, leading Rodriguez et al. (2003) to suggest that it is an MXBs. Its
nature has been confirmed by detection of pulsations in an {\it
  XMM-Newton} observation, with $P_{{\rm s}}=1300\:$s
(Lutovinov et al. 2004a). 

Pulsations have also been detected from a few other
absorbed {\it INTEGRAL} sources. IGR J16358$-$4726 displays pulsations
with a period of almost $6000\:$s, the second longest $P_{{\rm s}}$
known( Patel et al. 2004). Interestingly, these authors found that
IGR J16358$-$4726 appears to be transient, as it was
not detected during several {\it BeppoSAX} pointings of its field. As
many of the other new sources appear to be rather 
variable, it is likely that the $\sim10$ objects newly found represent
only a small fraction of the population of absorbed sources.

So far, five of the new {\it INTEGRAL} sources have been confirmed to
be X-ray pulsars. Based on the similarity between their hard spectra
and those of other absorbed {\it INTEGRAL} sources, Lutovinov et al. (2004b)
suggest that most of the new {\it INTEGRAL} sources are MXBs. This
seems to be confirmed by their obvious concentration toward the
Galactic arms (Lutovinov et al. 2004b).

In principle, the properties of these new sources do not appear to be
extremely different from those of known MXBs, except for the higher
absorption. The origin of this absorption is unclear. In the case of
IGR J16318$-$4848, it seems to be associated with heavy mass loss from
its peculiar companion. So far, the counterparts of most of these
sources are unknown.

However, there are reasons to think that some of the sources have
different characteristics from the three classes of MXB previously
known. The {\it INTEGRAL} source IGR J17391$-$3021 was detected during
two extremely short ($< 1\:$d) outbursts in August 2003, and promptly
identified with the already-known peculiar transient XTE~J1739$-$302,
which had displayed similar behaviour before (Smith et al. 1998).
Recent {\it Chandra} observations have allowed the identification of the
optical counterpart to XTE~J1739$-$302 (Smith et al. 2005), which turns
out to be an O8\,Iaf supergiant (Negueruela et al. 2005).

Shortly afterwards, a second source was detected by {\it INTEGRAL}
displaying the same sort of extremely short outbursts, IGR
J17544$-$2619. {\it XMM-Newton} observations allowed the detection of
a counterpart (Gonz\'alez-Riestra et al. 2004), which also turns out
to be an OB 
supergiant (Chaty et al. 2005). A third transient with very short
outbursts, IGR J16465$-$4507, discovered in September 2004, also has a
single candidate counterpart which could be an OB star
(Smith 2004).  IGR~J16465$-$4507 is certainly a MXBs, as pulsations
with $P_{{\rm s}}=228\:$s have been discovered in {\it XMM-Newton}
observations (Lutovinov et al. 2004b). Therefore there seems to be a new class
of MXBs  
characterised by very short outbursts. In at least two cases, the
counterparts are supergiants, clearly showing that not all MXBs with
supergiant companions share the characteristics traditionally
attributed to SXBs.

Some of the new {\it INTEGRAL} sources are likely to be similar to the
previously known population of MXBs. For example, the transient
IGR~J01363+6610 has a Be star as optical counterpart
(Reig et al. 2004b) and is almost certain to be a BeX
transient. However, the unprecedented characteristics of many of them,
such as the very high absorption the short outbursts or the transient
behaviour from supergiant systems rise the suspicion that we are
finding objects mostly belonging to new classes. The fact that many of
them appear transients opens the possibility that a large population of
MXBs remains still undetected.

\section{Conclusions}
 The traditional division of MXBs into three separate classes stems
 naturally from their physical characteristics and has indeed provided
 valuable insights that have allowed us to further our understanding
 of such systems. However, as new MXBs are being discovered, we are
 finding more and more systems that are basically impossible to fit
 into any of these divisions.

 On the one hand, the subclasses correspond
 to well-defined evolutionary stages, and some objects have been found
 that may be evolving toward or between some of these subclasses. On
 the other hand, the capability of {\it INTEGRAL} to detect sources
 strongly affected by photoelectric absorption results in the
 discovery of objects surrounded by denser environments. On both
 accounts, the discovery of new systems outside the classical
 classification appears only natural.

These discoveries, however, could have strong implications for our
understanding of MXBs. Until now, there has been the widespread
impression that the total Galactic population of SXBs could not be
very large. As these are persistent, relatively bright sources, they
can be seen to large distances (many of the known sources are at
distances of $\sim5\:$kpc) and hence we could expect the sample of
objects known to probe a significant fraction of the Galaxy. If many
new sources turn out now to be transient, the size of the population
could have been severely underestimated. 

We can expect {\it INTEGRAL} to discover several new MXBs in the
near future and {\it XMM-Newton} to provide error circles small enough
for the discovery of counterparts, allowing further characterisation
of this interesting class of objects.



\begin{acknowledgments}
The author is a researcher of the
programme {\em Ram\'on y Cajal}, funded by the Spanish MCyT
and the University of Alicante.
This research is partially supported by the Spanish MCyT under grants
AYA2002-00814 and ESP-2002-04124-C03-03. I would like to thank Marc
Rib\'o and Ingo Kreykenbohm for helpful comments on the draft.
\end{acknowledgments}





\begin{thebibliography}{zackerdackerf3}
\bibitem[Anzer \& B\"orner(1995)]{ab95}
Anzer, U., \& B\"orner, G. 1995, A\&A 299, 62
\bibitem[Baykal et al.(2000)]{bayk00}
Baykal, A., Stark, M.J., \& Swank, J. 2000, ApJ 544, L129
\bibitem[Campana et al.(2002)]{cam02}
Campana, S., et al. 2002, ApJ 580, 389
\bibitem[Chakrabarty et al.(1993)]{chak93}
Chakrabarty, D., et al. 1993, ApJ 403, L33
\bibitem[Chakrabarty et al.(2002)]{chak02}
Chakrabarty, D., et al. 2002, ApJ 573, 789
\bibitem[Charles et al.(1983)]{cha83} 
Charles, P.A., et al. 1983, MNRAS 202, 657
\bibitem[Chaty et al.(2005)]{chat05}
Chaty, S., et al. 2005, in preparation
\bibitem[Clark et al.(2002)]{clark02}
Clark, J.S., et~al. 2002, A\&A 392, 909
\bibitem[Corbet(1984)]{cor84} 
Corbet, R.H.D. 1984, A\&A 141, 91
\bibitem[Corbet(1986)]{cor86} 
Corbet, R.H.D. 1986, MNRAS 220, 1047
\bibitem[Corbet \& Peele(1997)]{cp97} 
Corbet, R.H.D., Peele, A.G. 1997, ApJ 489, L83
\bibitem[Delgado-Mart\'{\i} et al.(2001)]{delg01} 
Delgado-Mart\'{\i}, H., et al. 2001, ApJ 546, 455
\bibitem[Filliatre \& Chaty(2004)]{fc04}
Filliatre, P., \& Chaty, S. 2004, ApJ 616, 469
\bibitem[Finger et al.(1999)]{fin99}
Finger, M.H., et al. 1999, ApJ 517, 449
\bibitem[Galloway et al.(2004)]{gallo04}
Galloway, D.K., Morgan, E.H., \& Levine, A.M. 2004, ApJ 613, 1164
\bibitem[Gonz\'alez-Riestra et al.(2004)]{gonz04}
Gonz\'alez-Riestra, R., et al. 2004, A\&A 420, 589
\bibitem[Grimm et al.(2003)]{grimm03}
Grimm et al. 2003, MNRAS 339, 793
\bibitem[(Habets 1987)]{hab87} 
Habets, G.M.H.J. 1987, A\&A 184, 209
\bibitem[Hall et al.(2000)]{hall00}
Hall, T.A., et al. 2000, ApJ 536, 450
\bibitem[den Hartog et al.(2004)]{hartog04}
den Hartog, P.R., et al. 2004, ATel 281
\bibitem[Hayasaki \& Okazaki(2004)]{haya04}
Hayasaki, K., \& Okazaki, A.T. 2004, MNRAS 350, 971
\bibitem[van den Heuvel \& van Paradijs(1997)]{vhvp} 
van den Heuvel,  E.P.J., \& van Paradijs, J. 1997, ApJ 483, 339 
\bibitem[Hynes et al.(2002)]{hynes02}
Hynes, R.I., et al. 2002, A\&A 392, 991
\bibitem[Kaper(1998)]{kap98}
Kaper, L. 1998, 
In: I. D. Howarth (ed.), Proceedings Boulder-Munich Workshop II:
Properties of hot, 
luminous stars,  ASP Conf. Ser. 131, 427
\bibitem[King(1991)]{king91}
King, A. 1991, MNRAS 250, P3
\bibitem[Koh et al.(1997)]{koh97}
Koh, D.T., et al. 1997, ApJ 479, 933
\bibitem[Leahy(2002)]{leah02}
Leahy, D.A. 2002, A\&A 391, 219
\bibitem[Li \& van den Heuvel(1999)]{lvh99}
Li, X.-D., \& van den Heuvel, E.P.J. 1999, ApJ 513, L45
\bibitem[Liu et al.(2000)]{liu00} 
Liu, Q.Z., van Paradijs, J., \& van
  den Heuvel, E.P.J. 2000, A\&AS,147, 25 
\bibitem[Lutovinov et al.(2004a)]{lut04a}
Lutovinov, A., et al. 2004a, A\&A Letters, submitted ({\tt
  astro-ph/0411547})
\bibitem[Lutovinov et al.(2004b)]{lut04b}
Lutovinov, A., et al. 2004b, A\&A, submitted ({\tt astro-ph/0411550})
\bibitem[Majid et al.(2004)]{maj04}
Majid, W.A., Lamb, R.C., \& Macomb, D.J. 2004, ApJ 609, 133
\bibitem[Massi et al.(2004)]{massi04}
Massi, M., et al. 2004, A\&A 414, L1
\bibitem[Matsuda et al.(1992)]{mat92}
Matsuda, T., et al. 1992, MNRAS 255, 183
\bibitem[Matt \& Guainazzi(2003)]{mg03}
Matt, G., \& Guainazzi, M. 2003, MNRAS 341, L13
\bibitem[McGowen \& Charles(2003)]{mgc03}
McGowen, K.E., \& Charles, P.A. 2003, MNRAS 339, 748
\bibitem[Meynet \& Maeder(2003)]{mm03} Meynet, G., \& Maeder, A. 2003,
  A\&A  404, 975  
\bibitem[Negueruela(2005)]{neg05} 
Negueruela, I. 2005,
In: A.F.J. Moffat, \& N. St-Louis (eds.)
Massive stars in Interacting Binaries, ASP Conf. Series, in press
({\tt astro-ph/0411335}) 
\bibitem[Negueruela \& Coe(2002)]{nc02} 
Negueruela, I., \& Coe, M.J. 2002, A\&A 385, 517
\bibitem[Negueruela \& Reig(2001)]{nr01}
Negueruela, I., \& Reig, P. 2001, A\&A, 371, 1056
\bibitem[Negueruela \& Reig(2004)]{nr04}
Negueruela, I., \& Reig, P. 2004, ATel 285
\bibitem[Negueruela et al.(2001)]{neg01}
Negueruela, I., et al. 2001, A\&A 369, 117
\bibitem[Negueruela et al.(2003)]{neg03}
Negueruela, I., et al. 2003, A\&A 397, 739
\bibitem[Negueruela et al.(2005)]{nege05}
Negueruela, I., et al. 2005, in preparation
\bibitem[Okazaki(2001)]{oka01}
Okazaki, A.T. 2001, PASJ 53, 119
\bibitem[Okazaki \& Negueruela(2001)]{on01}
Okazaki, A.T., \& Negueruela, I. 2001, A\&A 377, 161
\bibitem[Okazaki et al.(2002)]{oka02}
Okazaki, A.T., et al. 2002, MNRAS 337, 967
\bibitem[Patel et al.(2004)]{patel04}
Patel, S.K., et al. 2004, ApJ 602, L45
\bibitem[Pfahl et al.(2002)]{pfahl02} 
Pfahl, E., et al. 2002, ApJ 574, 364
\bibitem[Podsiadlowski et al.(2004)]{pod04}
Podsiadlowski, Ph., et al. 2004, ApJ 612, 1044
\bibitem[(Portegies Zwart 1995)]{por95} 
Portegies Zwart, S.F. 1995, A\&A 296, 691 
\bibitem[Porter \& Rivinius(2003)]{pr03}
Porter, J.M., \& Rivinius, T. 2003, PASP 115, 1153
\bibitem[Quaintrell et al.(2003)]{quaint03}
Quaintrell, H., et al. 2003, A\&A 401, 313
\bibitem[Rappaport et al.(1978)]{rap78}
Rappaport, S., et al. 1978, ApJ 224, L1
\bibitem[Rappaport et al.(2004)]{rap04}
Rappaport, S., Fregeau, J.M., \& Spruit, H. 2004, ApJ 606, 436
\bibitem[Reig \& Roche(1999)]{rr99} 
Reig, P., \& Roche, P. 1999,  MNRAS 306, 100 
\bibitem[Reig et al.(2001)]{reig01}
Reig, P., et al. 2001, A\&A 367, 266
%
\bibitem[Reig et al.(2004a)]{reig04a}
Reig, P., et al. 2004a, A\&A 421, 673
\bibitem[Reig et al.(2004b)]{reig04b}
Reig, P., et al. 2004b, ATel 343
\bibitem[Reig et al.(2005)]{reig05}
Reig, P., et al. 2005, in preparation
\bibitem[Revnivtsev et al.(2003)]{rev03}
Revnivtsev, M., et al. 2003, AstL 29, 587
\bibitem[Rib\'o et al.(2002)]{ribo02}
Rib\'o, M., et al. 2002, A\&A 384, 954
\bibitem[Rodriguez et al.(2003)]{rod03}
Rodriguez, J., et al. 2003, A\&A 407, L41
\bibitem[Ruffert(1999)]{ruff99}
Ruffert, M. 1999, A\&A 346, 861
\bibitem[Ruffert \& Anzer(1995)]{ruff95}
Ruffert, M., \& Anzer, U. 1995, A\&A 295, 108
\bibitem[Schreier et al.(1972)]{sch72}
Schreier, E., et al. 1972, ApJ 172, L79
\bibitem[Smith et al.(1998)]{smith98}
Smith, D.M., et al. 1998, ApJ 501, L181
\bibitem[Smith(2004)]{smith04}
Smith, D.M., 2004, ATel 338
\bibitem[Smith et al.(2005)]{smith05}
Smith, D.M., et al. 2005, in preparation
\bibitem[Stella et al.(1986)]{swr86} 
Stella, L., White, N.E., \& Rosner, R. 1986, ApJ 308, 669 
\bibitem[Vink et al.(2000)]{vink00}
Vink, J.S., de Koter, A., \& Lamers, H.J.G.L.M. 2000, A\&A 362, 295
\bibitem[Waters \& van Kerkwijk(1989)]{wk89}
Waters, L.B.F.M., \& van  Kerkwijk, M.H. 1989, A\&A, 223, 196
\bibitem[Wilson et al.(2003)]{wil03}
Wilson, C.A., et al. 2003, ApJ 584, 996
\bibitem[Winkler et al.(2003)]{winkler03}
Winkler, C., et~al. 2003, A\&A 411, L1

\end{thebibliography}
\end{document}